\documentclass[aps,prl,onecolumn,showpacs]{revtex4}

\usepackage{multirow}
\usepackage{graphicx}
\usepackage{dcolumn}
\usepackage{amsmath}
\usepackage{amssymb}
\usepackage{wasysym}
\usepackage{color}
\usepackage{subfigure}
\begin{document}

\title[Quantum Oscillations in Iron Pnictide Superconductors]{Quantum Oscillations in Iron Pnictide Superconductors}

\author{Suchitra E. Sebastian}

\affiliation{Cavendish Laboratory, Cambridge University, JJ Thomson Avenue, Cambridge CB3~OHE, U.K.\\
suchitra@phy.cam.ac.uk}

\begin{abstract}
Insight into the electronic structure of the pnictide family of superconductors is obtained from quantum oscillation measurements. Here we review experimental quantum oscillation data that reveal a transformation from large quasi-two dimensional electron and hole cylinders in the paramagnetic overdoped members of the pnictide family to significantly smaller three-dimensional Fermi surface sections in the antiferromagnetic parent members, via a potential quantum critical point at which an effective mass enhancement is observed. Similarities with the Fermi surface evolution from the overdoped to the underdoped normal state of the cuprate superconducting family are discussed, along with the enhancement in antiferromagnetic correlations in both these classes of materials, and the potential implications for superconductivity.
\end{abstract}

\maketitle

\section{Quantum Oscillations}

Quantum oscillation measurements are a vital tool to characterise the electronic structure of metals. Originally conceived as a tool to probe elementary metals, quantum oscillations have grown in scope and have now been observed in complex intermetallics and oxides, including heavy fermions~\cite{Settai1,Sebastian0}, other unconventional superconductors~\cite{Bergemann1,Wosnitza1}, and even doped Mott insulating cuprate superconductors~\cite{Doiron1,Yelland1,Bangura1,Sebastian1,Sebastian2,Audouard1,Ramshaw1,Sebastian3a,Sebastian3}. The particular strength of quantum oscillations is their utility as a bulk probe of electronic structure, providing high momentum space resolution of constant energy surfaces at the Fermi energy.

\begin{figure}[htbp!]
\centering
\includegraphics*[width=.6\textwidth]{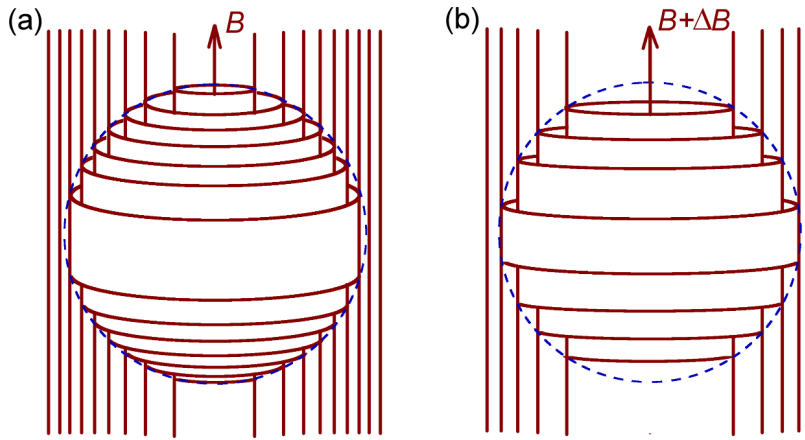}
\caption{(a) Schematic of Landau levels for spherical Fermi surface as in a free electron gas in a magnetic field $B$. (b) Schematic of Landau levels, when the applied magnetic field is increased to $B+\Delta B$. The Landau levels increase in area with an increase in magnetic field, causing fewer of them to be accommodated within the Fermi surface as the magnetic field is increased. [Modified from ref.~\cite{Shoenberg1}]}
\label{landau}
\end{figure}

In the presence of an applied magnetic field, a free electron gas whose surface of constant energy is spherical is arranged into a series of concentric cylinders known as `Landau tubes' (fig.~\ref{landau}). Electron motion is now confined to the surface of these `Landau tubes', resulting in quantised motion along `cyclotron orbits'. The quantisation of Landau tube areas arises from the quantum condition

\begin{eqnarray}
\Delta E = \hbar \omega_{\rm c}
\label{quantum}
\end{eqnarray}
\noindent where $\omega_{\rm c}$ is the angular frequency of an electron executing cyclotron motion, given by

\begin{eqnarray}
\omega_{\rm c}  = \frac{e B}{m^{\ast}} =\dfrac{2\pi e B/\hbar ^2}{\partial{\Lambda}/\partial{\epsilon}}
\label{omega}
\end{eqnarray}
\noindent where $m^{\ast}$ is the effective cyclotron mass, and $\Lambda$ is the cross sectional area in reciprocal space of a constant energy Landau tube intersected by a plane normal to the magnetic field $B$. From equations~\ref{quantum} and~\ref{omega}, the condition for Landau tube area quantisation is yielded as

\begin{eqnarray}
\Lambda_r=(r+\frac{1}{2})\frac{2\pi e B}{\hbar}
\label{landaueqn}
\end{eqnarray}
\noindent When the magnetic field is increased, each Landau tube expands in area (shown in fig.~\ref{landau}), causing its height accommodated within the Fermi surface to shrink, until it spans the extremity of the Fermi surface (with extremal area $\Lambda=A$) just before it `pops' out of the Fermi surface entirely. As each Landau tube pops out of the Fermi surface, there is a sudden discontinuous change in the total energy of the occupied states. Such an event happens periodically as tubes of successively small quantum numbers $r$ pass through the Fermi surface at equal intervals of $1/B$ (from eqn.~\ref{landaueqn}). The frequency of Landau tubes popping out is given by

\begin{eqnarray}
F=\frac{\phi_0 A}{2\pi^2} = 2\frac{A}{A_{\rm BZ}} \frac{\phi_0}{a b}
\label{onsager}
\end{eqnarray}
\noindent where $\phi_0=\frac{h}{e}$ is the magnetic flux quantum, $A_{\rm {BZ}}$ is the area of the Brillouin zone in reciprocal space, and $a$ and $b$ are the unit cell dimensions in real space assuming a tetragonal lattice~\cite{Shoenberg1}. We label the frequency harmonics as $pF$, where $p$ represents the $p$th harmonic. All derivatives of the free energy, such as magnetisation, consequently display oscillatory behaviour. These quantum oscillations in magnetisation are referred to as `de Haas$-$van Alphen' oscillations, and in electrical transport are referred to as `Shubnikov de Haas' oscillations~\cite{Shoenberg1}. The frequency of oscillations reflects the area of the extremal Fermi surface orbit. A complex Fermi surface geometry with multiple orbits, or a Fermi surface comprising multiple sections would yield an oscillatory signal comprising multiple frequencies, which can be analysed by filtering or Fourier transform methods.

\subsection{Angular dependence}

\subsubsection{Fermi surface geometry}

Quantum oscillations have an important role in characterising Fermi surface geometry, with the potential to reveal significant information about the physics of the systems studied, especially in strongly correlated systems. For instance, in heavy fermion systems, the geometry of the Fermi surface can reveal information about the participation or otherwise of $f$-electrons in the electronic conduction of the system~\cite{Settai1,Sebastian0}. In underdoped high temperature cuprate superconductors, quantum oscillations measurements of the Fermi surface geometry have proved an important indicator of likely Fermi surface reconstruction by a long range ordering wavevector~\cite{Audouard1,Ramshaw1,Sebastian3}.

The area of extremal orbits of the Fermi surface can be studied with great precision from the measured quantum oscillation frequencies. To further identify the momentum space location of the Fermi surface sections and map out the precise Fermi surface geometry, comparison with calculated theoretical band structures is typically required. Measurement of the quantum oscillation frequency as a function of the magnetic field direction specifies the extremal areas of the Fermi surface normal to all directions relative to the crystalline axes. This provides geometrical information about the Fermi surface that can be compared with band structure calculations. For a quasi-two dimensional Fermi surface, the angular dependence provides information relating to the $c$-axis dispersion of the Fermi surface~\cite{Yamaji1}.

\subsubsection{Spin splitting}
The quantum oscillation amplitude as a function of angle is also modulated by an effect known as spin splitting: separate sets of Landau levels arise from spin-up and spin-down electrons, leading to a phase difference between oscillations from spin-up and spin-down electrons, and hence a modulation of the oscillation amplitude due to interference. Whereas the Zeeman splitting between spin up and spin down electrons is invariant with angle in the absence of significant spin-orbit effects, the Landau level separation is a function of the projected magnetic field $B \cos \theta$ for quasi-two dimensional systems, where $\theta$ is the angle between the applied magnetic field and the crystalline axis. The quantum oscillation amplitude as a function of angle is consequently modulated by a prefactor

\begin{eqnarray}
R_{\rm s}=\cos \big(\frac{1}{2}\frac{\pi g_\theta^\ast m_\theta^\ast}{m_{\rm e}})
\end{eqnarray}
\noindent where $g^\ast_\theta$ is the effective spin-splitting factor, and $m^\ast_\theta$ is the effective quasiparticle mass corresponding to a magnetic field applied at an angle $\theta$ to the crystalline $c$-axis. In cases where the spins are ordered in such a way that the spin moment is suppressed, the interference term is not expected to appear $-$ potentially proving a diagnostic tool in detecting forms of spin order~\cite{Ramshaw1,Sebastian3}.

\section{Magnetic field dependence}
An important source of quantum oscillation amplitude modulation is `phase smearing' (i.e.) effects occur that are equivalent to the superposition of oscillations over a finite phase window, resulting in an amplitude modulation of the net oscillations. One such amplitude reduction factor ($R_{\rm Dingle}$, denoted by $R_{\rm D}$) arises from a finite energy spread due to quasiparticle scattering time, and is given by
\begin{eqnarray}
R_{\rm D}=e^{\frac{-B_{\rm c}}{B}}
\label{Dingle}
\end{eqnarray}
\noindent where  $\frac{B_{\rm c}}{B}= \frac{p\pi}{\omega_{\rm c}\tau}=\frac{p\pi r_{\rm c}}{l}$; $\hbar\omega_{\rm c}$ represents the Landau level spacing, $\tau$ is the scattering time, $l$ is the mean free path, and $r_{\rm c}$ is the cyclotron radius. As a consequence, crystal quality is paramount in observing sufficiently large quantum oscillations. The exponential rate of growth of the experimentally observed quantum oscillation amplitude with magnetic field yields a measure of the purity of the measured single crystal, with a faster rate of growth indicating a higher impurity concentration.

\section{Temperature Dependence}
Another source of phase smearing is finite temperature. For quasiparticles with Fermi Dirac statistics, the temperature dependence of the quantum oscillation amplitude is governed by the sharp Fermi Dirac distribution that grows more rounded with increasing temperature. A reduction at finite temperatures in the abruptness with which the particle occupation as a function of energy vanishes causes phase smearing, and a consequent reduction in the amplitude of the measured quantum oscillations. In this case, the magnitude of amplitude reduction as a function of temperature is given by a derivative of the Fermi Dirac distribution
\begin{eqnarray}
R_{\rm T}=\frac{\pi \lambda}{\sinh \pi \lambda}
\label{tempdamp}
\end{eqnarray}
\noindent where $\lambda = \frac{2 p \pi m^\ast k_{\rm B} T}{e \hbar B}$, $k_{\rm B}$ is the Boltzmann constant. For $\pi \lambda \gtrapprox 1$,

\begin{eqnarray*}
	R_T \approx 2\pi\lambda e^{-\pi \lambda}=\frac{4 \pi^2pk_{\rm B}T}{\hbar \omega_{\rm c}}e^{-\frac{2\pi^2pk_{\rm B}T}{\hbar\omega_{\rm c}}}
\end{eqnarray*}
\noindent where $\lambda$ is expressed in terms of the Landau level spacing $\hbar\omega_{\rm c}$, analogous to the magnetic field dependent damping. A measure of quantum oscillation amplitude as a function of temperature is used to directly verify Fermi Dirac statistics~\cite{Sebastian1}, and to measure the effective cyclotron mass of elementary quasiparticles.

The enhancement in effective quasiparticle mass over the band mass is a direct measure of the strength of correlations, of particular importance in strongly correlated materials. A value of effective cyclotron mass $m^\ast$ is extracted by a fit to the temperature dependence of the experimentally measured quantum oscillation amplitude. The effective mass can furthermore, serve as a diagnostic of a quantum critical point at which a relevant susceptibility diverges. Quantum oscillations have proven an effective way to detect a quantum critical point, especially underlying an unconventional superconducting phase by measuring a divergent effective mass (i.e. a collapse in Fermi velocity) on the suppression of superconductivity by an applied magnetic field~\cite{Settai1,Sebastian5}. Furthermore, the vicinity of a quantum critical point may be a propitious location in phase space to look for a breakdown of Fermi Dirac statistics, given that a divergence in effective mass signals a breakdown in the quasiparticle concept~\cite{Coleman1}.

\section{Iron Pnictide Superconductors}

\begin{figure}[htbp!]
\centering
\includegraphics*[width=0.75\textwidth]{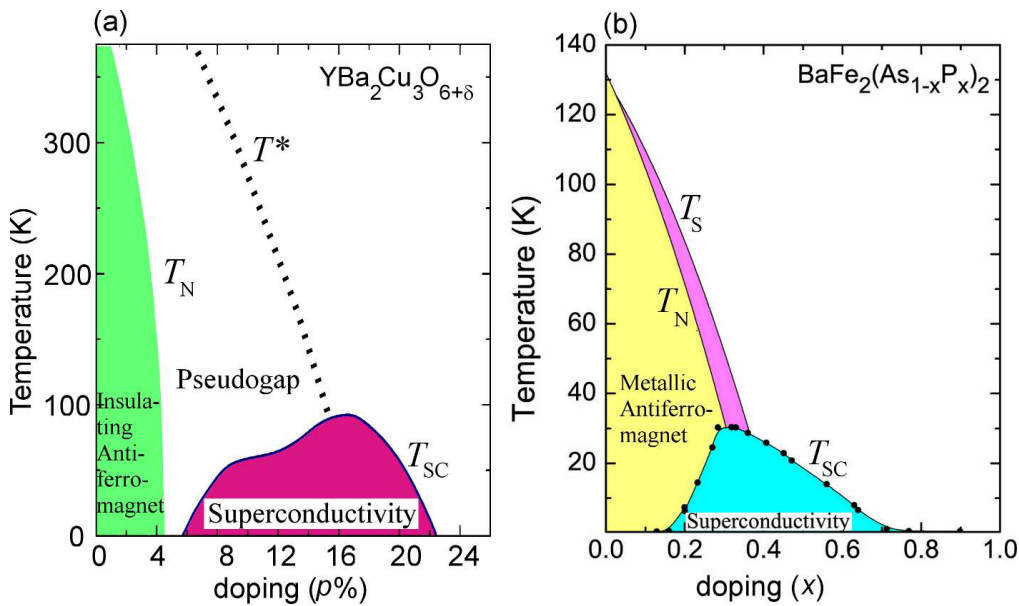}
\caption{(a) Schematic phase diagram for a copper oxide high temperature superconductor, the example of YBa$_2$Cu$_3$O$_{6+\delta}$ is shown in the figure (for hole doping $p \approx 18 \delta \%$), from ref.~\cite{Liang1}. At zero to low dopings, an insulating antiferromagnetic phase with N$\acute{e}$el temperature $T_{\textrm N}$ is exhibited, evolving with higher doping to a $d$-wave superconductor with superconducting temperature $T_{\rm sc}$ and a `pseudogap' normal state that onsets at $T^\ast$. (b) Schematic phase diagram for an iron pnictide superconductor (the example of BeFe$_2$(As$_{1-\textrm{x}}$P$_{\rm{x}}$)$_2$ from ref.~\cite{Jiang1,Kasahara1} shown here). At zero to low dopings, a metallic antiferromagnetic phase with N$\acute{e}$el temperature $T_{\rm N}$ preceded by a structural phase transition at $T_{\rm s}$ is exhibited, evolving at higher dopings to a superconductor with superconducting temperature $T_{\rm sc}$.}
\label{phasediag}
\end{figure}

We now turn to the iron pnictide family of superconductors, and the measurements of quantum oscillations in these materials. A landmark advance in the field of superconductivity was made with the discovery of the family of superconducting iron pnictides~\cite{Kamihara1,Kamihara2}. These tetragonal structure-type materials that contain tetrahedrally coordinated Fe$^{2+}$ on a square lattice superconduct at temperatures as high as 56~K~\cite{Ren1}. Proving next only to the copper oxide superconductors in superconducting temperatures, iron pnictide materials are of particular interest due to the unconventional nature of superconductivity they exhibit~\cite{Norman1}. Materials include the `1111' family with the ZrCuSiAs-structure type, containing $R$FeAsO members ($R$=rare earth)~\cite{Kamihara1,Kamihara2,Ren1}, the `122' family with the ThCr$_2$Si$_2$-structure type, containing $A$Fe$_2$As$_2$ ($A$=alkaline earth, alkali metal)~\cite{Rotter1,Krellner1,Dong1} members, the `111' family with the Ce$_2$Sb-structure type~\cite{Wang1,Tapp1,Pitcher1}, containing $A$FeAs members, and the `11' family with the PbO-structure type, containing Fe$Ch$ ($Ch$=chalcogen)~\cite{Hsu1} members, among others. Nearly all these materials such as the `1111' and `122' families are antiferromagnetic in their groundstate~\cite{Rotter1,Krellner1,Cruz1}, and become superconducting on the application of either charge or isovalent chemical doping~\cite{Kamihara2,Ren1,Rotter2,Jiang1,Kobayashi1,Kasahara2,Sefat1,Kasahara1} or the application of pressure~\cite{Alireza1,Mizuguchi1}. Intriguingly, the phase diagram of the pnictides has some broad similarities with other unconventional superconductors such as heavy fermions, organics, and high temperature cuprate superconductors~\cite{Norman1} (shown in figure~\ref{phasediag}).

\section{Quantum oscillations in antiferromagnetic parent iron pnictides}

\begin{figure}[htbp!]
\centering
\includegraphics*[width=.7\textwidth]{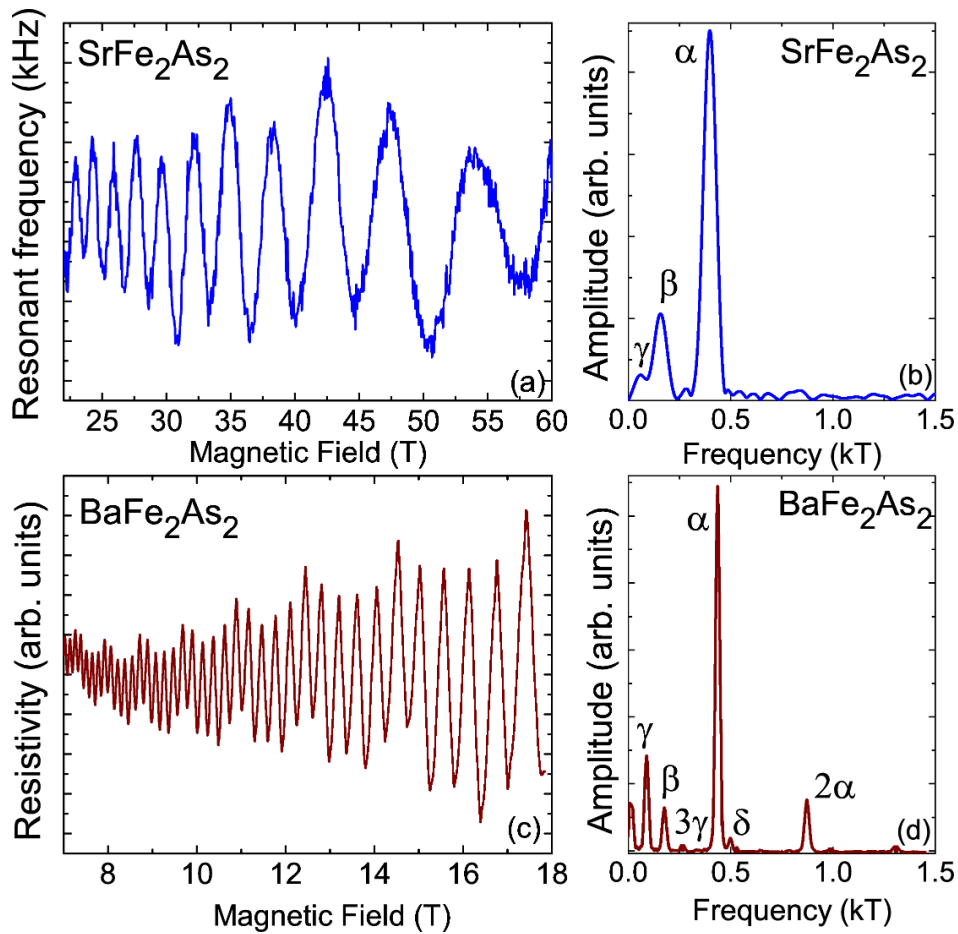}
\caption{(a) Quantum oscillations measured by the tunnel diode oscillator technique on a single crystal of SrFe$_2$As$_2$ between 22~T and 60~T at an applied magnetic field angle of 10$^\circ$ to the crystalline $c$-axis and a temperature of 0.5~K. (b) Fourier transform of the quantum oscillations in SrFe$_2$As$_2$ shown in (a), with peaks identifying quantum oscillation frequencies that correspond to $F_{\alpha}$=370(20)~T, $F_{\beta}$=140(20)~T, and $F_{\gamma}$=70(20)~T for an applied magnetic field along the crystalline $c$-axis. From~ref.~\cite{Sebastian4,Sutherland1}. (c) Quantum oscillations measured in the resistivity of a single detwinned crystal of BaFe$_2$As$_2$ from 7~T to 18~T, with the magnetic field parallel to the crystalline $c$-axis, at a temperature of 0.17~K. (d) Fourier transform of the quantum oscillations in BaFe$_2$As$_2$ shown in (c), with peaks identifying quantum oscillation frequencies that correspond to $F_{\alpha}$=440~T, $F_{\beta}$=170~T, $F_{\gamma}$=90~T, and $F_{\delta}$=500~T, for an applied magnetic field along the crystalline $c$-axis. From~ref.~\cite{Terashima1}.
}
\label{AFoscns}
\end{figure}

Soon after the discovery of superconductivity in the iron pnictide family of materials, quantum oscillations were discovered in the antiferromagnetic parent iron arsenide SrFe$_2$As$_2$ using the resonant oscillatory technique in a pulsed field magnet and torque magnetometry in a DC hybrid magnet~\cite{Sebastian4}, and the superconducting iron phosphide LaFePO  using torque and resistivity techniques in superconducting and DC hybrid magnets~\cite{Coldea2}. Quantum oscillations have now been measured in other members of the pnictide family of superconductors including other `122' and `111' materials by a variety of experimental techniques including magnetic torque measurements, resonant oscillatory measurements, and electrical transport measurements in a range of magnets - from pulsed, to DC resistive and hybrid, to superconducting magnets~\cite{Analytis1,Harrison1,Terashima1,Sutherland1,Harrison2,Analytis2,Coldea1,Arnold1,Analytis3,Terashima2,Putzke1,Coldea2}. Here we focus on quantum oscillations in the parent and overdoped `122' family and the evolution of quantum oscillations as a function of doping in these materials.

\begin{figure}[htbp!]
\centering
\includegraphics*[width=.7\textwidth]{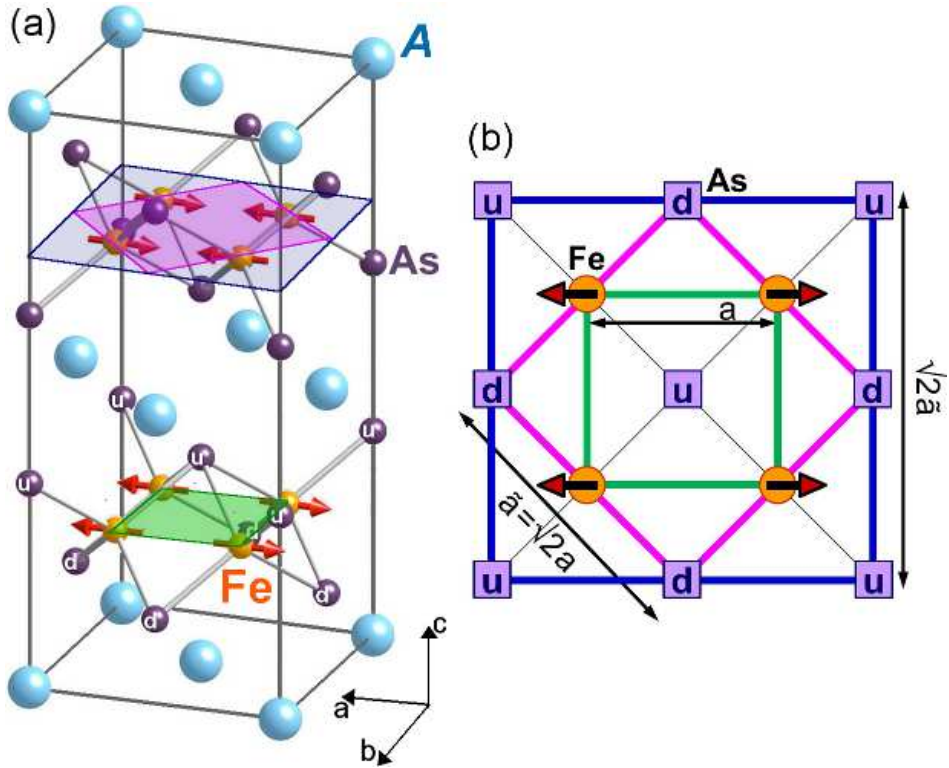}
\caption{(a) Unit cell of the antiferromagnetic `122' family of $A$Fe$_2$As$_2$ pnictides: $A$=Sr,Ba,Ca. $A$ atoms shown in blue, Fe ions in orange, As ions in purple. Ordering of Fe moments (shown in red) arranged parallel to each other along the $b$ direction, and antiparallel toward each other along the orthogonal $a$ direction. The one-Fe ion unit cell is shown in green, with side $a$ (top-view shown in (b)). However, the arrangement of As ions located above (labelled `u') and below (labelled `d') the plane of Fe-ions causes the minimal unit cell to be defined as a larger unit cell containing two Fe ions, shown in pink, with side $\tilde{a}=\sqrt{2}a$ (top-view shown in b). The antiferromagnetic unit cell is obtained by doubling the two-Fe ion unit cell, and is shown in blue. The side of the antiferromagnetic unit cell is given by $\sqrt{2}a$, the top view is shown in b. Modified from ref.~\cite{Goldman1}. (b) top view of each of the one-Fe ion, two-Fe ion, and doubled antiferromagnetic unit cells accompanied by dimensions, and the alignment of Fe-spins due to antiferromagnetic order.
}
\label{unitcell}
\end{figure}

\begin{figure}[htbp!]
\centering
\includegraphics*[width=.7\textwidth]{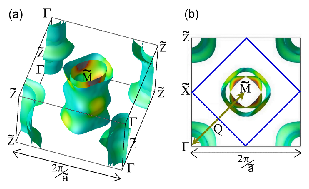}
\caption{(a) Calculated paramagnetic Fermi surface of SrFe$_2$As$_2$ in the original nonmagnetic two-Fe ion Brillouin zone, calculated using using the local density approximation (LDA) method, comprising two hole cylinders located at the $\Gamma$ point and two electron cylinders at the $\tilde{M}$ point of the Brillouin zone. High symmetry points in the two-Fe ion Brillouin zone are denoted as $\Gamma$, $\tilde{M}$, $\tilde{X}$, $\tilde{Z}$. Shading of the surfaces is according to the calculated Fermi velocity. From ref.~\cite{Sebastian4}. (b) A top down view of the nonmagnetic two-Fe ion unit cell accompanied by the antiferromagnetic ordering wavevector $Q=\big(\pi/\tilde{a},\pi/\tilde{a} \big)$~\cite{Cruz1} to yield an antiferromagnetic unit cell (shown in blue). From~ref.~\cite{Sebastian1}.}
\label{AF_folding}
\end{figure}

Figures~\ref{AFoscns}a,c show a sample of quantum oscillations measured over different magnetic field ranges in single crystals of SrFe$_2$As$_2$ and detwinned single crystals of BaFe$_2$As$_2$, which are antiferromagnetic parent pnictides with ordering temperatures of 205~K and 140~K respectively~\cite{Krellner1,Rotter1}. Prominent oscillations periodic in inverse magnetic field are measured in the field range 7~T to 60~T. Figures~\ref{AFoscns}b,d show Fourier transforms of measured quantum oscillations in SrFe$_2$As$_2$ and BaFe$_2$As$_2$ over different field ranges. Low frequency Fourier peaks are observed at $F_\alpha$=370(20) T, $F_\beta$=140(20) T, $F_\gamma$=70(20) T in SrFe$_2$As$_2$~\cite{Sebastian4}, and at $F_\delta$=500 T, $F_\alpha$=440 T, $F_\beta$=170 T, $F_\gamma$=90 T in BaFe$_2$As$_2$~\cite{Terashima1,Analytis1}.

\subsection{Fermi surface geometry: nonmagnetic and antiferromagnetic bandstructure calculations}

\begin{figure}[htbp!]
\centering
\includegraphics*[width=.6\textwidth]{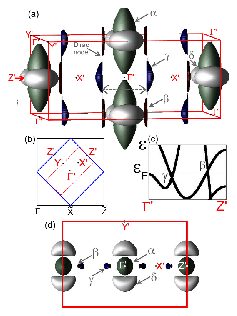}
\caption{(a) Calculated antiferromagnetic Fermi surface of SrFe$_2$As$_2$ obtained using the local density approximation (LDA) method in which magnetic ordering is included~\cite{Analytis1,Harrison2}. High symmetry points in the new antiferromagnetic Brillouin zone are denoted by $\Gamma'$, $X'$, $Y'$, $Z'$. (b) The relative dimensions of the reduced antiferromagnetic Brillouin zone (in red) are compared to the conventional original nonmagnetic two-Fe ion (in black). While the $\alpha$, and $\delta$ surfaces denote the larger ellipsoidal surfaces formed from parabolic bands, the $\gamma$ and $\beta$ pockets are smaller closed surfaces with pointed ends formed from linear band crossings at a Dirac point (shown in (c)). The dashed arrow shows a sample construction of the $\beta$ and $\gamma$ Dirac pockets at a Dirac point. (d) shows the top view of the Brillouin zone with Fermi surface pockets and high symmetry points labelled. A comparison of the experimentally measured extremal orbits with the band structure calculations is shown in Table~\ref{tableunder}. Figures provided courtesy of M. D. Johannes. 
}
\label{AF_Dirac}
\end{figure}

The crystal structure of the `122' iron pnictide is shown in figure~\ref{unitcell}a. Shown in green in figures~\ref{unitcell}a,b is the basal plane unit cell containing a single Fe ion, of side $a$. However, the alternating positioning of the As ions above (marked as `u') and below (marked as `d') the Fe-ion plane (shown in figures~\ref{unitcell}a,b) mean that the primitive unit cell contains two Fe ions. The basal plane unit cell containing two Fe ions of side $\tilde{a}$ (where $\tilde{a}=\sqrt{2}a$) is shown in pink in figures~\ref{unitcell}a,b. For the remainder of this chapter, we adopt this two-Fe ion notation in describing the unit cell and the Brillouin zone, and the ordering wavevector within the Brillouin zone. An alternate convention arises on noting that the low energy part of the electronic structure is equivalent to a Brillouin zone that is twice as large, and corresponds to a unit cell $axa$ in dimension, containing a single Fe ion. The first convention can be related to the second, by noting that the $X=\big(\pi/a,0\big)$ and $Y=\big(0,\pi/a\big)$ points in the one-Fe unit cell correspond to the $\tilde{M}=\big(\pi/\tilde{a},\pi/\tilde{a}\big)$ point in the two-Fe unit cell. In the nonmagnetic state, local density approximation (LDA) calculations yield an electronic structure of $A$Fe$_2$As$_2$ that comprises two electron-like surfaces near the $\tilde{M}$ point, and two hole surfaces near the $\Gamma$ point (shown in fig.~\ref{AF_folding})~\cite{Sebastian4,Singh1}.

It is clear that the experimentally measured Fermi surface sections, each occupying $\lessapprox 2 \%$ of nonmagnetic Brillouin zone ($\%$ of Brillouin zone area occupied by each Fermi surface section shown in table~\ref{tableunder}) are vastly smaller in size than the quasi-two-dimensional sections occupying up to $\approx 25 \%$ of the nonmagnetic Brillouin zone expected in the calculated nonmagnetic bandstructure. A Fermi surface transformation into sections of much smaller size is expected from reconstruction due to an additional superstructure, which would increase the size of the unit cell in real space, thereby reducing the size of the Brillouin zone in momentum space, and causing the Fermi surface to break up into smaller sections. Such a superstructure in antiferromagnetic $A$Fe$_2$As$_2$ might be expected to occur at the quasi-nesting wavevector $Q=\big(\pi/\tilde{a},\pi/\tilde{a}\big)$ between the electron and hole pockets which are separated in momentum space in the nonmagnetic electronic structure, given the enhanced Lindhard function in the vicinity of this wavevector~\cite{Mazin1,Kuroki1,Chubukov1,Cvetkovic1}. Indeed, the onset of an antiferromagnetic wavevector $Q=\big(\pi/\tilde{a},\pi/\tilde{a}\big)$ at the N$\acute{e}$el temperature has been measured by inelastic neutron scattering in the parent pnictide materials, in which neighbouring Fe moments are arranged parallel to each other along one direction, and antiparallel to each other along the orthogonal direction (shown in figure~\ref{unitcell})~\cite{Cruz1}. In SrFe$_2$As$_2$ and BaFe$_2$As$_2$, a spin moment of $\approx~0.9~\mu_{\rm B}$ per Fe ion is measured~\cite{Zhao1,Huang1}, and indications from the breadth of the measured high energy dispersion are that the spin density wave is itinerant in character~\cite{Ewings1}.

\begin{table}[ht!]
\caption{Measured quantum oscillation frequencies ($F$) in antiferromagnetic parent `122' pnictide materials SrFe$_2$As$_2$ and BaFe$_2$As$_2$, corresponding Fermi surface cross sectional areas as a fraction of the original two-Fe ion Brillouin zone area ($\frac{A}{A_{\rm {BZ}}}$), measured effective quasiparticle mass ($\frac{m^\ast}{m_{\rm e}}$), effective quasiparticle mass enhancement compared to the band mass ($\frac{m^\ast}{m_{\rm b}}$). The corresponding Fermi surface sections are illustrated in figure~\ref{AF_Dirac}. From refs.~\cite{Sebastian4,Analytis1,Terashima1}.}
{\begin{tabular}
{|l|l|l|l|l|l|l|l|l|}
\hline
&\multicolumn{4}{|c|}{SrFe$_2$As$_2$}&\multicolumn{4}{|c|}{BaFe$_2$As$_2$}\\
\hline
Fermi surface&$F (\rm T)$&$\frac{A}{A_{\rm {BZ}}}(\%)$&$\frac{m^{\ast}}{m_{\rm e}}$&$\frac{m^{\ast}}{m_{\rm B}}$&$F (\rm T)$&$\frac{A}{A_{\rm {BZ}}}(\%)$&$\frac{m^{\ast}}{m_{\rm e}}$&$\frac{m^{\ast}}{m_{\rm B}}$\\
\hline
${\alpha}$ (hole)&$370(20)$&$1.38$&2.0(1)&$1.8$&$440$&$1.7$&2.1(1)&$2.8$\\
\hline
${\beta}$ (Dirac)&$140(20)$&$0.52$&1.5(1)&$2.1$&$170$&$0.7$&1.8&$-$\\
\hline
${\gamma}$ (Dirac)&$70(20)$&$0.26$&$-$&$-$&$90$&$0.3$&0.9&$-$\\
\hline
${\delta}$ (electron)&$-$&$-$&$-$&$-$&$500$&$1.9$&2.4(3)&$2.0$\\
\hline
\end{tabular}
}
\label{tableunder}
\end{table}

The anticipated electronic structure in this parent antiferromagnetic state as a consequence of the $Q=\big(\pi/\tilde{a},\pi/\tilde{a}\big)$ ordering wavevector is yielded by folding the non-magnetic Brillouin zone shown in Figure~\ref{AF_folding}a along the magnetic wavevector $Q$, such that the $\tilde{M}$ point is folded onto the $\Gamma$ point. The unit cell in the basal plane is therefore doubled, yielding the antiferromagnetic unit cell shown in dark blue in figures~\ref{unitcell}a,b. The resulting Brillouin zone is reduced in size from the black square shown in figure~\ref{AF_folding}a,b to the blue diamond shown in figure~\ref{AF_folding}b. The reduction in size of the Brillouin zone results in the superposition of the hole pocket at $\Gamma$, and the electron pocket at $\tilde{M}$, causing a gapping of large sections of the original Fermi surface, and leaving behind small three-dimensional hole and electron pockets~(shown in figure~\ref{AF_Dirac})~\cite{Sebastian4}. Figure~\ref{AF_Dirac} shows the result of a local density approximation calculation of the antiferromagnetic state of SrFe$_2$As$_2$ in which magnetic ordering is included; a value of negative on-site Coulomb repulsion $U$ is used that suppresses the magnetic moment to the experimentally observed value of $\approx~0.9~\mu_{\rm B}$~\cite{Harrison2,Analytis1}. The calculated antiferromagnetic Fermi surface (figure~\ref{AF_Dirac}a,d) is seen to comprise small three dimensional Fermi surface pockets in contrast to the larger quasi-two dimensional Fermi surface sections expected in the nonmagnetic electronic structure (figure~\ref{AF_folding}).

\subsection{Experimental comparison with bandstructure}

\begin{figure}[htbp!]
\centering
\includegraphics*[width=.5\textwidth]{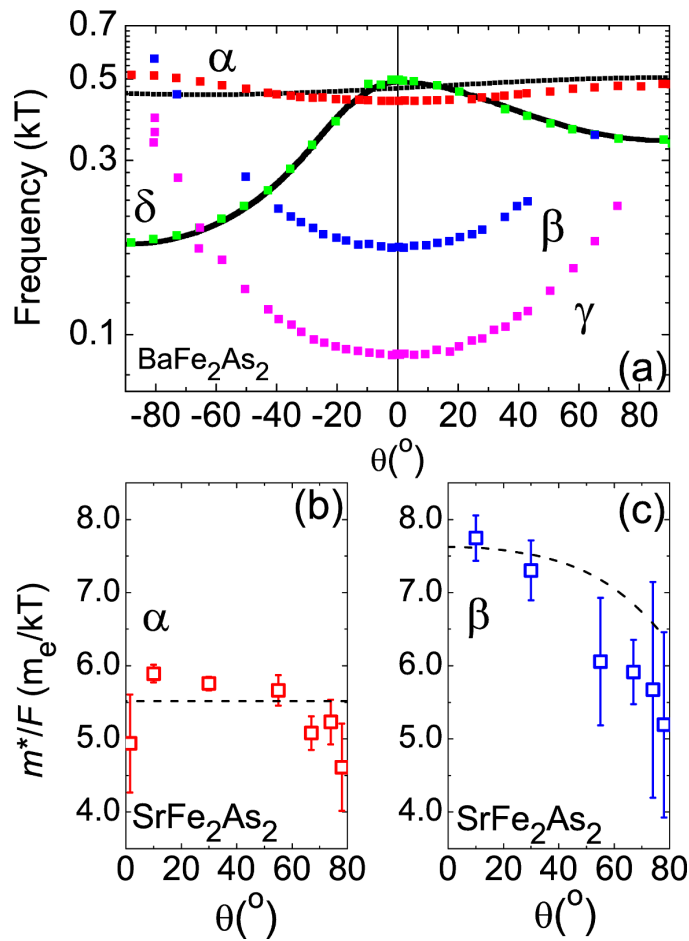}
\caption{(a) Experimentally measured angular dependence of the $\alpha$, $\beta$, $\gamma$, and $\delta$ frequencies in BeFe$_2$As$_2$ compared with expected angular dependence of $\alpha$ and $\delta$ from a calculated Fermi surface (similar to that shown in figure~\ref{AF_Dirac}). Good agreement is seen for the $\alpha$ and $\delta$ frequencies for a band shift less than 65 meV, corresponding to a magnetic moment size of 1.65 $\mu_B$ per Fe ion. From ref.~\cite{Terashima1}. (b) Angular dependence of the ratio of effective mass to quantum oscillation frequency experimentally measured for the $\alpha$ pocket in SrFe$_2$As$_2$ - the ratio is largely constant (dashed line is a guide to the eye), as expected for a parabolic band (c) angular dependence of the ratio of effective mass to frequency experimentally measured for the $\beta$ pocket in SrFe$_2$As$_2$ - the value dips sharply for higher angles (dashed line is a guide to the eye), consistent with the expectation for linear bands crossing at a Dirac point as described in section~\ref{Dirac}. Modified from ref.~\cite{Sutherland1}.
}
\label{AFangular}
\end{figure}

Table~\ref{tableunder} shows the measured quantum oscillation frequencies and effective quasiparticle masses in the antiferromagnetic parent pnictide SrFe$_2$As$_2$ and BaFe$_2$As$_2$, along with the effective mass enhancement of each Fermi surface section, and fraction of the original two-Fe ion Brillouin zone area occupied by each Fermi surface section. The size and angular dependence of the measured quantum oscillation frequencies that correspond to the extremal orbits of the Fermi surface pockets are compared with expectations from the calculated antiferromagnetic bandstructure in figure~\ref{AFangular}. Figure~\ref{AFangular}a shows a comparison of the angular dependence of extremal orbits expected from the calculated band structure, with an angular dependence of the measured quantum oscillation frequencies in BaFe$_2$As$_2$~\cite{Terashima1}. The observed frequencies extend over the entire angular range, and are weakly angle-dependent, indicating largely three-dimensional Fermi surface sections. Good agreement between experimental and calculated frequencies is seen for the $\alpha$ and $\delta$ frequencies for a band shift of -44meV and +65meV respectively for the bands corresponding to the larger electron-like $\delta$ pocket and hole-like $\alpha$ pocket shown in figure~\ref{AF_Dirac}. Interestingly, this band shift corresponds to a calculated magnetic moment of 1.6$\mu_B$ per Fe ion, which overestimates the experimentally measured magnetic moment of 0.9 $\mu_B$ per Fe ion, potentially indicating the relevance of fully magnetic models in describing the parent antiferromagnetic pnictide materials.

The identification of the $\gamma$ and $\beta$ pocket is less unambiguous, given their extremely small size, as seen in figure~\ref{AF_Dirac}. A more detailed study of these small Fermi surface pockets is presented in section~\ref{Dirac}. Using the observed enhancement in effective mass compared to the calculated band mass for the Fermi surface pockets measured in SrFe$_2$As$_2$ and BaFe$_2$As$_2$ (table~\ref{tableunder}), the contribution to the Sommerfeld coefficient from the measured pockets is estimated to be 3.5 mJ/K$^2$mol for SrFe$_2$As2$_2$ and 5.0 mJ/K$^2$mol for BaFe$_2$As$_2$, in good agreement with measured values of heat capacity~\cite{Sebastian4,Analytis1,Terashima1}. The modest effective mass enhancement places these materials in a regime of intermediate correlation strength. Quantum oscillation measurements of the Fermi surface in the parent $A$Fe$_2$As$_2$ family enables experimental identification of the reconstructed Fermi surface as corresponding to the calculated antiferromagnetic bandstructure in this material, reconstructed from the original bandstructure by an ordering wavevector $Q=\big(\pi/\tilde{a},\pi/\tilde{a}\big)$.

\subsection{Dirac Nodes}\label{Dirac}

A more in-depth comparison is required to identify the topology of the smallest $\gamma$ and $\beta$ pockets observed in the Fermi surface of antiferromagnetic SrFe$_2$As$_2$ and BaFe$_2$As$_2$. A theoretical analysis shows that the experimentally observed spin density wave gives rise to symmetry-enforced band degeneracy at high-symmetry points in the Brillouin zone, the experimentally observed spin density wave with ordering wavevector $Q=\big(\pi/\tilde{a},\pi/\tilde{a} \big)$ resulting in a nontrivial band topology. Even were nesting to be perfect, topological protection would preserve the Dirac nodes that result for such a band topology, thereby preventing full gapping of the Fermi surface by the spin density wave; a conclusion that has been shown to survive the introduction of strong interactions~\cite{Ran1}. The location of Dirac nodes which are located close to the Fermi energy in the case of antiferromagnetic SrFe$_2$As$_2$ and BaFe$_2$As$_2$~\cite{Analytis1,Sutherland1,Harrison2} (shown in figure~\ref{AF_Dirac}) yields small Dirac pockets. The geometry of the small closed Dirac pockets with pointed ends is not sufficient to distinguish these from ellipsoidal pockets; a true test of such a Dirac point can only be yielded by a comparison of the angular dependence of the associated frequency and effective cyclotron mass, in this case, for the $\beta$ and $\gamma$ pockets. For a parabolic band, this ratio would remain constant; for a linear crossing, however, the angular-dependent ratio of the frequency and effective cyclotron mass is expected to deviate from a constant value. For a linear dispersion of the form $\epsilon = \pm \hbar v^\ast |k|+ 2t\cos(c k_z / 2) + \mu$ where $c$ is the bilayer spacing for the body-centred-tetragonal crystal structure (12.3 \AA ), $v^\ast$ is the characteristic Dirac velocity, $t$ is the interlayer hopping parameter, $|k|= \sqrt{k_x^2+k_y^2}$, and $\mu$ is the chemical potential, a striking experimental signature is expected. As $\theta$ is increased, the orbits would begin to encompass regions of the Fermi surface close to the Dirac point (fig.~\ref{AF_Dirac} left inset), where velocities are high and masses are very light - leading to a reduction in orbitally averaged cyclotron masses, and consequently a rapid reduction in $m^\ast / F$ with angle compared to the expectation for a parabolic band~\cite{Harrison1,Sutherland1}.

Figures~\ref{AFangular}b,c show the ratio of $m^\ast / F$ as a function of angle for both the $\alpha$ pocket and the $\beta$ pocket in SrFe$_2$As$_2$. The ratio $m^\ast / F$ remains largely constant as a function of angle for the $\alpha$ pocket, consistent with its parabolic band origin. However, the ratio $m^\ast / F$ dips steeply with angle in the case of the $\beta$ pocket, as would be expected for a pocket associated with linear band crossing at a Dirac point~\cite{Sutherland1}. The linear dispersion characteristic of a Dirac cone is also directly observed via photoemission experiments~\cite{Richard1}. The Fermi surface of the antiferromagnetic parent `122' pnictide family is therefore seen to exhibit Dirac points as a consequence of the orbital character of the bands in this material~\cite{Sutherland1,Harrison2,Ran1}.

\section{Quantum oscillations in overdoped paramagnetic iron pnictides}

\begin{figure}[htbp!]
\centering
\includegraphics*[width=0.7\textwidth]{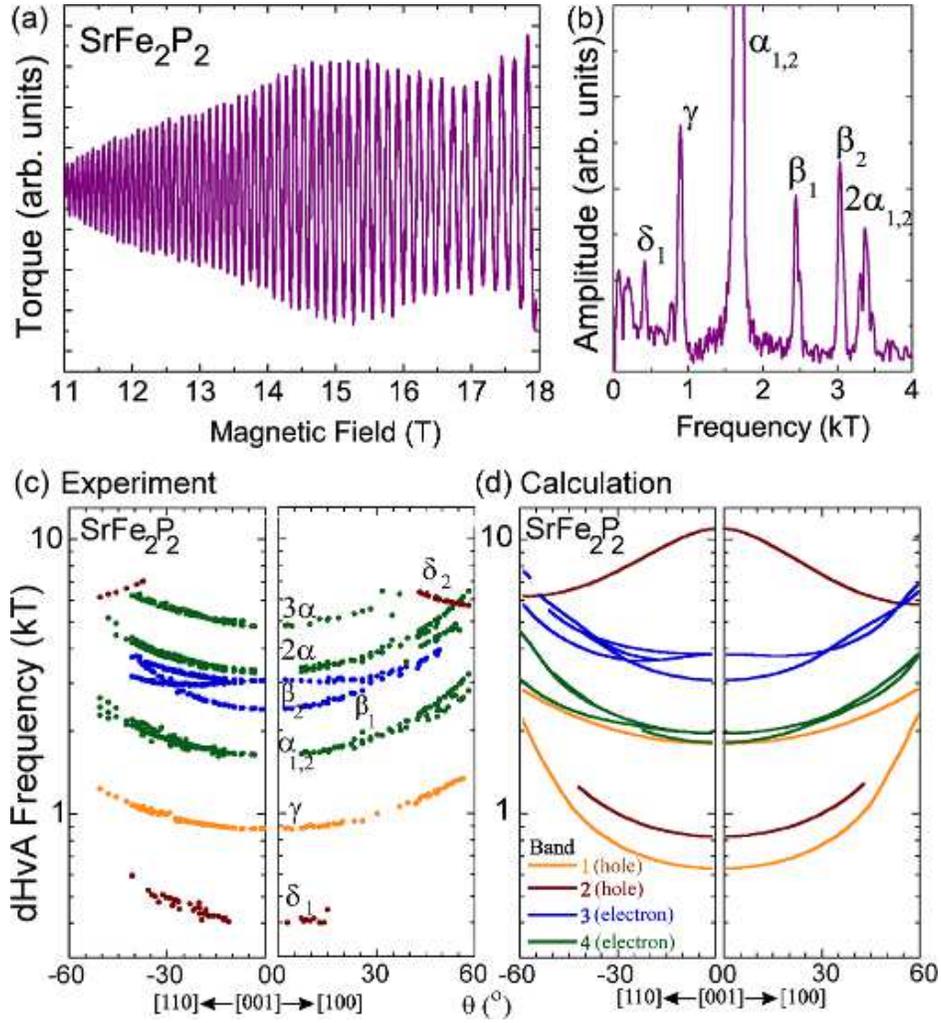}
\caption{(a) Quantum oscillations measured using torque magnetometry on a single crystal of SrFe$_2$P$_2$ (at an angle of 9$^\circ$ to the crystalline $c$-axis) (b) Fourier analysis of the quantum oscillations shown in (a) to extract the constituent frequencies from Fourier peaks labelled according to the Fermi surface sections to which they correspond. (c) Angular dependence of experimentally measured quantum oscillation frequencies. (d) Calculated angular dependence of quantum oscillation frequencies expected for the band structure shown in figure~\ref{122Fermisurfaces}. Excellent agreement is seen between the angular dependence of the experimentally measured quantum oscillations and extremal orbits calculated from the band structure. From ref.~\cite{Analytis2}.
}
\label{SrFe2P2}
\end{figure}

Quantum oscillations have been observed in the end $A$Fe$_2$P$_2$ members of the superconducting `122' pnictide family (fig.~\ref{phasediag})~\cite{Analytis2,Coldea1,Arnold1}. Figure~\ref{SrFe2P2} shows quantum oscillations measured on SrFe$_2$P$_2$~\cite{Analytis2}. The first noticeable aspect of these oscillations is that the measured frequencies are considerably higher than in the parent antiferromagnetic parent member of the series (fig.~\ref{AFoscns}), indicating larger Fermi surface sections (measured frequencies and the size of the corresponding Fermi surface cross sectional area as a fraction of the original two-Fe ion Brillouin zone are shown in table~\ref{tableover}). The Fourier transform in figure~\ref{SrFe2P2}b shows the component frequencies in the measured quantum oscillations, the angular dependence of which is shown in figure~\ref{SrFe2P2}c. A band structure calculation of the expected Fermi surface of SrFe$_2$P$_2$ consists of two concentric electron cylinders located at the Brillouin zone corners (i.e. the $\tilde{M}$ point), and two hole cylinders centred at the $\Gamma$ point (figure~\ref{122Fermisurfaces}), which is very similar to that calculated for the nonmagnetic phase of SrFe$_2$As$_2$ (figure~\ref{AF_folding}a). Some band structure calculations~\cite{Xu1} and photoemission experiments~\cite{Brouet1} experiments indicate an additional third hole pocket located at the $\Gamma$ point, that is more three-dimensional in character.

Measured quantum oscillation frequencies and effective mass enhancements are shown in table~\ref{tableover} for SrFe$_2$P$_2$ and BaFe$_2$P$_2$. These yield experimental results that correspond well with band structure calculations, with the largest frequency corresponding to a cross-sectional area as large as $\approx 25 \%$ the size of the original two-Fe ion Brillouin zone. Figures~\ref{SrFe2P2}c,d show the angular dependence of the experimentally measured quantum oscillation frequencies compared with the predictions for the calculated band structure in SrFe$_2$P$_2$. Good agreement is obtained on shifting electron bands $\alpha$ and $\beta$ up by 59 and 49 meV respectively, and the hole band $\gamma$ down by 110 meV~\cite{Analytis2} in the calculated band structure. Using the observed enhancement in effective mass compared to the calculated band mass for the Fermi surface pockets measured in SrFe$_2$P$_2$, the contribution to the Sommerfeld coefficient from the measured Fermi surface pockets is estimated to be 10.4(2) mJ/mol/K$^2$, in good agreement with the value of 11.6(2) mJ/mol/K$^2$ measured by heat capacity. The size of quasiparticle effective mass enhancement places these materials in a regime of intermediate correlation strength. The significant transformation in Fermi surface from nonmagnetic end members of the `122' series to the antiferromagnetic parent members of this family of materials is discussed more in Section~\ref{qcpsection}.

\subsection{Quasi-nesting of hole and electron cylinders}

\begin{figure}[htbp!]
\centering
\includegraphics*[width=.7\textwidth]{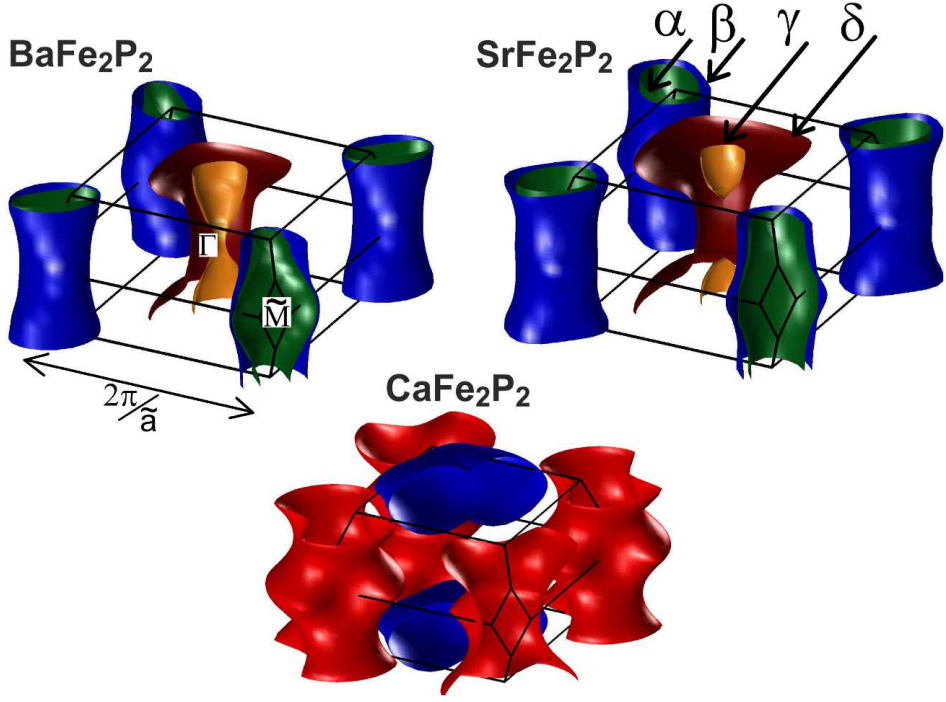}
\caption{Fermi surfaces of BaFe$_2$P$_2$, SrFe$_2$P$_2$, and CaFe$_2$P$_2$ from Local Density Approximation calculations with bands shifted for best agreement with experimental quantum oscillation data~\cite{Analytis2, Coldea1, Arnold1}. Bands corresponding to extremal orbits $\alpha$, $\beta$, $\gamma$, and $\delta$ indicated by arrows. The quasi-two dimensional cylinders in SrFe$_2$As$_2$ and BaFe$_2$As$_2$ are warped, meaning that surfaces $\alpha$, $\beta$, and $\delta$ each have a minimal (orbit$_1$) and maximal (orbit$_2$) extremal orbit corresponding to the `neck' and `belly' of the warped surface. A comparison of the experimentally measured extremal orbits with the band structure calculations is shown in Table~\ref{tableover}. From ref.~\cite{Carrington1}.}
\label{122Fermisurfaces}
\end{figure}

An important aspect of the Fermi surface geometry revealed by quantum oscillations in the $A$Fe$_2$P$_2$ end members of the `122' family is the significant $c-$axis dispersion (i.e. warping) of all the Fermi surface sections, rendering them more three dimensional instead of two dimensional. It is also obvious that while the Lindhard function (representing the bare spin susceptibility) is expected to have a maximum at $Q=\big(\pi/\tilde{a},\pi/\tilde{a}\big)$, a perfect nesting condition is not satisfied by the pocket geometry. Nesting effects are far from ideal in either the unreconstructed band structure of the antiferromagnetic parent members~(figure~\ref{AF_folding}), or the measured Fermi surface of the end members of the pnictide family~(figure~\ref{122Fermisurfaces}), making it unlikely that these effects are principally responsible for antiferromagnetism.

\begin{table}[htbp!]
\caption {Measured quantum oscillation frequencies $F$ in the end members SrFe$_2$P$_2$ and BaFe$_2$P$_2$ of the `122' pnictide family, corresponding Fermi surface cross-sectional areas as a fraction of the original two-Fe ion Brillouin zone area ($\frac{A}{A_{\rm {BZ}}}$), measured effective quasiparticle mass ($\frac{m^\ast}{m_{\rm e}}$), effective quasiparticle mass enhancement compared to the band mass ($\frac{m^\ast}{m_{\rm b}}$). The corresponding Fermi surface sections are illustrated in figure~\ref{122Fermisurfaces}. From refs.~\cite{Analytis2,Arnold1}}
{\begin{tabular}
{|l|l|l|l|l|l|l|l|l|}
\hline
&\multicolumn{4}{|c|}{SrFe$_2$P$_2$}&\multicolumn{4}{|c|}{BaFe$_2$P$_2$}\\
\hline
Fermi surface&$F (\rm kT)$&$\frac{A}{A_{\rm {BZ}}}(\%)$&$\frac{m^{\ast}}{m_{\rm e}}$&$\frac{m^{\ast}}{m_{\rm B}}$&$F (\rm kT)$&$\frac{A}{A_{\rm {BZ}}}(\%)$&$\frac{m^{\ast}}{m_{\rm e}}$&$\frac{m^{\ast}}{m_{\rm B}}$\\
\hline
${\gamma}$ (hole)&$0.89$&$3.4$&1.49(2)&$1.4$&$1.25$&$4.8$&1.64&$1.80$\\
\hline
${\delta_1}$ (hole)&$0.41$&$1.6$&1.6(1)&$1.3$&$0.79$&$3.1$&2.30&$1.59$\\
\hline
${\delta_2}$ (hole)&$6.02$&$23.2$&3.41(5)&$1.7$&$6.58$&$25.4$&3.32&$1.59$\\
\hline
${\beta_1}$ (electron)&$2.41$&$9.3$&1.92(2)&$1.6$&$2.14$&$8.3$&1.65&$1.68$\\
\hline
${\beta_2}$ (electron)&$3.06$&$12.0$&2.41(3)&$1.6$&$2.31$&$8.9$&1.54&$1.81$\\
\hline
${\alpha_1}$ (electron)&$1.637$&$6.3$&$1.13(1)$&$2.1$&$1.18$&$4.8$&1.75&$1.82$\\
\hline
${\alpha_2}$ (electron)&$1.671$&$6.5$&$1.13(1)$&$2.1$&$1.35$&$4.6$&1.58&$1.88$\\
\hline
\label{tableover}
\end{tabular}
}
\end{table}

Quantum oscillation measurements have been performed on the other end members of the `122' pnictide family: BaFe$_2$P$_2$, and CaFe$_2$P$_2$~\cite{Coldea1,Arnold1,Analytis3}. Figure \ref{122Fermisurfaces} shows the calculated Fermi surface geometry corresponding to quantum oscillations measured in each of these families of materials~\cite{Analytis2,Coldea1,Arnold1,Carrington1}. On traversing the periodic table from the Ba members of the series, to the Sr and Ca members of the series, an interesting evolution is seen. The pocket sizes and geometries in SrFe$_2$P$_2$ and BaFe$_2$P$_2$ are fairly similar, with the chief difference being the better fulfilment of nesting criteria between inner hole ($\gamma$) and outer electron ($\alpha$) pockets in BaFe$_2$P$_2$.  The Fermi surface sections in CaFe$_2$P$_2$, however, become particularly different, with the outer electron pockets becoming more warped, and the central warped concentric hole cylinders in BaFe$_2$P$_2$ and SrFe$_2$P$_2$ transforming into a single three dimensional surface in CaFe$_2$P$_2$ (figure~\ref{122Fermisurfaces}). In fact, the electronic structure in CaFe$_2$P$_2$ corresponds to that of the collapsed tetragonal structure, where the ratio of $c$-axis to $a$-axis lattice constant is significantly reduced from the orthorhombic parent antiferromagnetic structure~\cite{Kasahara2}. A potential relation of the increased quasi-three dimensionality in CaFe$_2$P$_2$ to the reduced optimal-$T_{\rm {sc}}$ in this families of materials (optimal $T_{\rm {sc}}$ $\approx 30~$K in BaFe$_2$(As$_{\rm{1-x}}$P$_{\rm x}$)$_2$ and SrFe$_2$(As$_{\rm{1-x}}$P$_{\rm x}$)$_2$, compared to optimal $T_{\rm {sc}}$ $\approx 15~$ K in CaFe$_2$(As$_{\rm{1-x}}$P$_{\rm x}$)$_2$) is consequently indicated~\cite{Kasahara1,Kobayashi1,Kasahara2}.

\section{Cuprates and iron pnictides - electronic structure comparison}

\begin{figure}[htbp!]
\centering
\includegraphics*[width=.7\textwidth]{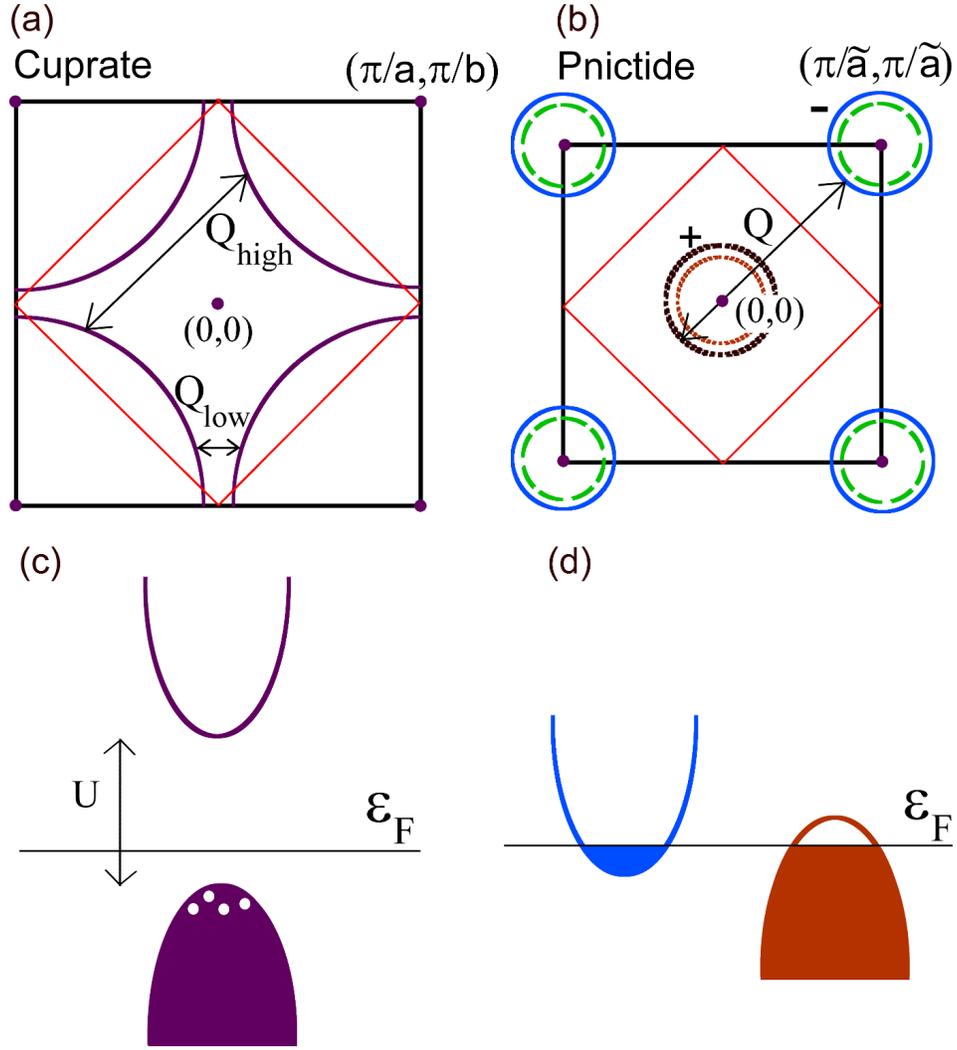}
\caption{(a) Schematic of single hole band in the original Brillouin zone of a hole-doped cuprate superconductor. $Q_{\rm high}=\big(\pi/\tilde{a},\pi/\tilde{a}\big)$ and $Q_{\rm low}=\big(\Delta,0 \big)$ show intra-Fermi surface nesting wavevectors at which the Lindhard function is enhanced. The red diamond indicates the antiferromagnetic Brillouin zone. (b) Schematic of the hole pockets (indicated by $`+'$) and electron pockets (indicated by $`-'$) from multibands in the original Brillouin zone of a pnictide superconductor. $Q=\big(\pi/\tilde{a},\pi/\tilde{a}\big)$ shows the nesting wavevector between hole and electron pockets. The red diamond indicates the antiferromagnetic Brillouin zone. (c) Schematic of the half-filled band split by onsite Coulomb repulsion $U$ to yield a Mott insulator with filled lower Hubbard band, and empty upper Hubbard band in the parent cuprate materials. The white dots show doped hole carriers that constitute the single large hole Fermi surface shown in (a). (d) Schematic of a compensated semimetal representing the pnictide materials with electron and hole bands separated in momentum space, yielding multiple Fermi surfaces as shown in (b).
}
\label{bands}
\end{figure}

Importantly, quantum oscillations are a key tool in making a comparison between the normal state of the cuprate and pnictide superconductors. At first glance, the cuprate and pnictide phase diagrams appear similar, both manifesting an antiferromagnetic parent groundstate that evolves to a superconducting state with elevated superconducting temperatures on doping~(figure~\ref{phasediag}). However, the metallic antiferromagnetic groundstate of the pnictide superconductors persists with doping, overlapping with the superconducting region over a significant portion of the phase diagram, up to optimal dopings. In contrast, the antiferromagnetic Mott insulating groundstate of the cuprates persists over a narrow range with doping, and is separated in doping from the onset of the superconducting dome. A key question pertains to the nature of the normal state of the cuprates in the underdoped region of superconductivity. Low frequency quantum oscillations were discovered after two decades in the underdoped region of the phase diagram in the cuprate superconductor YBa$_2$Cu$_3$O$_{6+{\delta}}$ at hole dopings $p$ $\approx 9\% - 12\%$ (i.e. ${\delta} = 0.49 - 0.61$), corresponding to a Fermi surface of area $\approx 2 \%$ of the original Brillouin zone~\cite{Doiron1,Sebastian1,Sebastian2,Ramshaw1,Sebastian3a,Sebastian3,Audouard1}. Quantum oscillations have subsequently been observed in other underdoped cuprate superconductors, including YBa$_2$Cu$_4$O$_8$~\cite{Yelland1,Bangura1} and Nd$_{\rm{2-x}}$Ce$_{\rm x}$CuO$_4$~\cite{Helm1}, revealing similarly small Fermi surface sections.

\subsection{Enhancement in Lindhard function in pnictides and cuprates}

An interesting comparison can be drawn between the Fermi surface in the cuprates and the pnictides with possible relevance to the origin of superconductivity in these materials~\cite{Basov1}. At first sight, the nonmagnetic Fermi surface in the cuprate materials appears very different from the pnictide materials. The pnictide family of materials belong to the class of compensated semimetals, with an electronic structure comprising both hole and electron pockets $-$ these are separated in momentum space, as seen in figures~\ref{AF_folding},\ref{122Fermisurfaces}. In contrast, the cuprate materials are effectively single band in character. A significant Jahn-Teller distortion in the quasi-two dimensional cuprates removes the degeneracy of the two-fold $e_g$ orbital in copper, and leads to an effective single orbital model. The intra-orbital Coulomb interaction $U$ causes the half filled band to be split into a filled lower Hubbard and an empty upper Hubbard band (fig.~\ref{bands}c), yielding a Mott insulating ground state. On doping mobile carriers as shown in figure~\ref{bands}c, a single carrier band is yielded. The electronic structure on the overdoped hole-carrier side has been measured by angular magnetoresistance oscillations, quantum oscillations, and angle resolved photoemission, revealing a single hole surface centred at the $\Gamma$ point of the Brillouin zone (schematic shown in figure~\ref{bands}a)~\cite{Vignolle1,Sebastian3}. A single hole band occupying $\approx 65 \%$ of the original Brillouin zone has been observed by quantum oscillations on the overdoped side of the hole-doped cuprate superconductor Tl$_2$Ba$_2$CuO$_{6+{\delta}}$~\cite{Vignolle1}.

A notable commonality, however, is seen between the electronic structure of the cuprate and pnictide families on identifying the quasi-nesting wavevector associated with a peak in the Lindhard function in both these cases. In the case of the pnictide materials, a quasi-nesting wavevector $Q=\big(\pi/\tilde{a},\pi/\tilde{a}\big)$ connects the electron pockets at the $\Gamma$ location with the hole pockets at the $M$ location in the pnictides, manifesting itself as a peak in the Lindhard function~\cite{Mazin1,Kuroki1,Chubukov1,Cvetkovic1,Mazin2}. The system then becomes unstable to an excitonic instability (i.e.) a combination of electrons and holes with relative momentum wavevector $Q$, experimentally found to be manifested as an antiferromagnetic spin density wave~\cite{Halperin1}. Indeed, this ordering wavevector at which the Lindhard function is maximised is responsible for Fermi surface reconstruction into the observed small Fermi surface sections in the parent members of the pnictide family.

In the case of the cuprates, an intraband wavevector $Q_{\rm {high}} = \big(\pi/\tilde{a},\pi/\tilde{a}\big)$ connecting opposite sides of the single large band and associated with antiferromagnetic correlations is found to enhance the Lindhard function~\cite{Monthoux1}. In this case, the Lindhard function is also enhanced at small wavevectors such as $Q_{\rm {low}} = (\Delta,0)$, which is associated with charge density wave type of order, and has been related to Fermi surface reconstruction into the small Fermi surface sections observed by quantum oscillations in the underdoped regime~\cite{Sebastian3}. The $Q=\big(\pi/\tilde{a},\pi/\tilde{a}\big)$ wavevector associated with antiferromagnetic correlations at which the Lindhard function is enhanced in both pnictide and cuprate families of materials could contribute to the enhancement of superconducting temperatures in these and other unconventional superconductors~\cite{Lonzarich1,Norman1,Basov1}. A possible test of the relevance of antiferromagnetic correlations in contributing to elevated superconducting temperatures is in the case of multiband systems where the location of electron and hole pockets coincide in momentum space; here the contribution of antiferromagnetic correlations to superconductivity would be expected to be suppressed.

\subsection{Quantum critical point under superconducting dome}\label{qcpsection}

\begin{figure}[htbp!]
\centering
\includegraphics*[width=.7\textwidth]{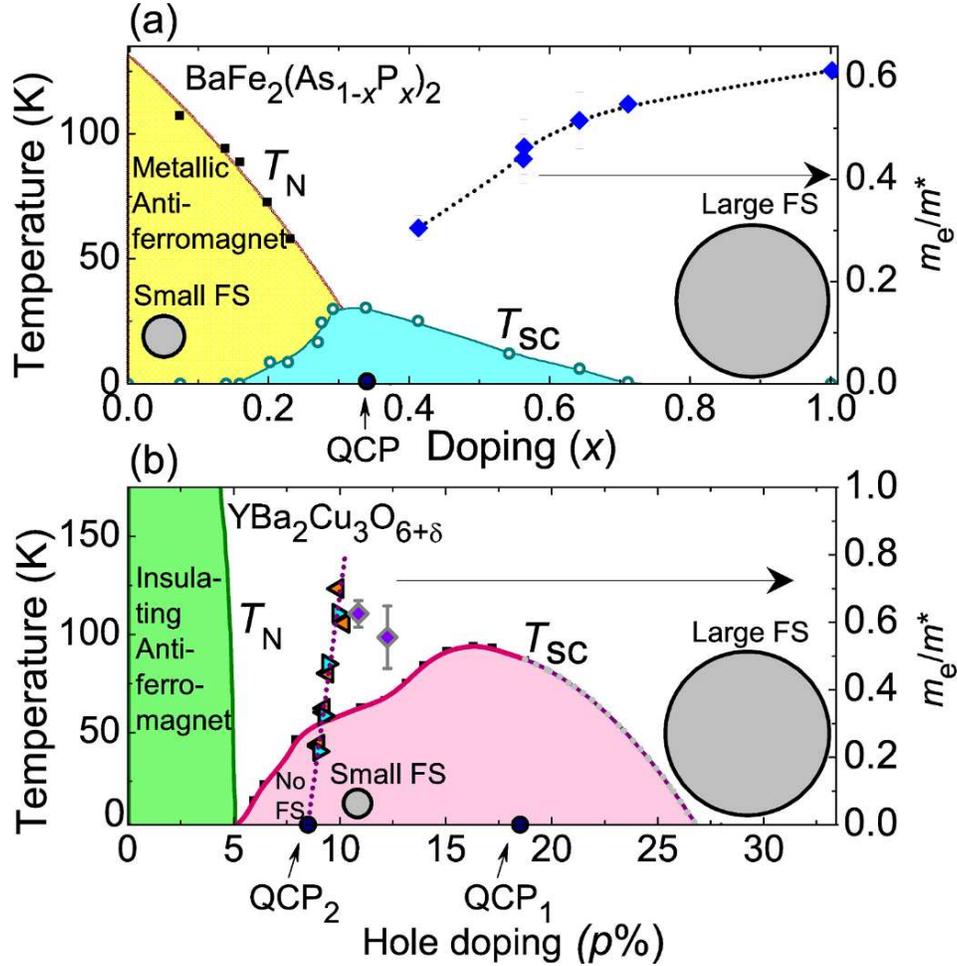}
\caption{(a) Phase diagram of the doped pnictide BaFe$_2$(As$_{1-\rm x}$P$_{\rm x}$), showing the drastic change in Fermi surface from a small area in the undoped metallic antiferromagnetic regime to a large area in the overdoped region (illustrated by gray circles with area proportional to the largest measured quantum oscillation frequency). A putative quantum critical point (labelled QCP) at the antiferromagnetic instability at the end of the metallic antiferromagnetic phase, which separates the small from the large Fermi surface is indicated by the collapse of experimentally measured Fermi velocity (proportional to the inverse effective quasiparticle mass shown on the rhs axis) on approaching optimal doping. Modified from ref.~\cite{Shishido1}. (b) Phase diagram of the high temperature cuprate superconductor YBa$_2$Cu$_3$O$_{6+{\delta}}$. In this case the antiferromagnetic phase on the undoped side is insulating. A quantum critical point is indicated by a collapse in the experimentally measured Fermi velocity (proportional to the inverse quasiparticle mass shown on the rhs axis) at a low doping denoted as QCP$_2$, likely separating a charge density wave phase above QCP$_2$ in which the Fermi surface area is small, and a spin density wave phase below QCP$_2$ in which no Fermi surface is observed~\cite{Sebastian3}. Left-facing and right-facing triangle symbols are from quantum oscillations measured up to 55~T and 85~T respectively, from ref.~\cite{Sebastian5}; diamond symbols are from ref.~\cite{Singleton1}. A notional quantum critical point (QCP$_1$) is expected at optimal doping separating the underdoped region in which a small Fermi surface is observed, and the overdoped region in which a large Fermi surface is observed in the Tl$_2$Ba$_2$CuO$_{6+\delta}$ family of materials~\cite{Vignolle1}. Relative Fermi surface areas are illustrated by gray circles with area proportional to the measured quantum oscillation frequency. Modified from ref.~\cite{Sebastian5}.
}
\label{qcp}
\end{figure}

Models involving electronic mediation have been suggested to explain unconventional superconductivity in materials such as the cuprates and pnictides. In these models, an enhancement in electronic susceptibility, such as for example the density or magnetic susceptibility is anticipated to favour superconductivity~\cite{Lonzarich1}. While qualitative insight may be gained from the Lindhard function which represents the bare spin susceptibility, an experimentally determined electronic susceptibility is the best indication of where maximally enhanced superconductivity may be expected within such models. A magnetic instability manifested as a quantum critical point at zero temperature would yield an enhanced spin susceptibility. An attendant challenge, however, pertains to experimentally accessing the vicinity of a quantum critical point, given that a new phase such as superconductivity often conceals the notional quantum critical point. Quantum oscillations provide a good experimental probe in the vicinity of a quantum critical point, at which the effective quasiparticle mass would be expected to diverge.

In the case of pnictide superconductors, superconductivity can be suppressed sufficiently by applied magnetic fields up to 65~T, such that quantum oscillations are observed in the doping range $x=0, 0.41-1$ in the BaFe$_2$(As$_{1-\rm x}$P$_{\rm x}$)$_2$ family~\cite{Shishido1}. The doping range accessed by quantum oscillations extends very close to optimal doping and reaching dopings at which the superconducting temperatures are as high as $T_{\rm {sc}} \approx 25$~K ($\approx 0.8 T_{\rm {sc}}^{\rm {max}}$). Intriguingly, an increase in effective mass, corresponding to a fall in Fermi velocity is observed in the region approaching optimal doping~(shown in figure~\ref{qcp}a). The Fermi surface geometry remains largely unchanged above optimal doping, corresponding to quasi-nested warped cylindrical electron and hole pockets at the $\tilde{M}$ and $\Gamma$ points of the Brillouin zone respectively, which reduce in size by $\approx 30 \%$ from $x=1$ to $x=0.41$~\cite{Shishido1}. A drastic change is, however, observed compared to the Fermi surface geometry in the undoped regime, in which the largest Fermi surface extremal area is approximately five times smaller than the largest Fermi surface extremal area for $x=1$. The decrease in Fermi velocity between $x=1$ and $x=0.41$ in BaFe$_2$(As$_{1-\rm x}$P$_{\rm x}$)$_2$ is consistent with an ultimate collapse of Fermi velocity (proportional to the inverse effective quasiparticle mass) at a quantum critical point located at the doping where the antiferromagnetic phase boundary would end, were superconductivity to be absent. The collapse in Fermi velocity approaching optimal doping is consistent with a divergent magnetic susceptibility at an antiferromagnetic instability marking the notional quantum critical point underlying the superconducting maximum.

In the case of cuprate superconductors, a drastic change is observed between the Fermi surface extremal area in the overdoped regime and in the underdoped regime, where the extremal area is less than thirty times smaller than in the overdoped regime. However, the use of quantum oscillations to detect evidence for a divergent effective mass at optimal doping separating these two regimes has proved challenging given the robustness of superconductivity in this region of the phase diagram, and the high magnetic fields required to suppress superconductivity. Intriguingly, quantum oscillations measured up to 85 T observe a steep increase in effective cyclotron mass, corresponding to a collapse in Fermi velocity approaching a lower doping (hole doping $p \approx 8.5 \%$ (i.e. $\delta \approx 0.45$) in YBa$_2$Cu$_3$O$_{6+{\delta}}$) in the vicinity of a plateau (local maxima) in the superconducting dome (figure~\ref{qcp}b)~\cite{Sebastian5}. In this case, the nature of the enhanced susceptibility associated with the quantum critical point signalled by the rapid collapse in Fermi velocity has not yet been precisely ascertained, but is likely associated with the onset of a spin density wave below this doping, and charge density wave above this doping~\cite{Sebastian3}.

In both these families of cuprate and pnictide superconductors, and in other families of unconventional superconductors~\cite{Norman1}, similar quantum critical points located at a magnetic or other form of instability underlying the superconducting dome maximum appear consistent with common mechanisms of unconventional superconductivity mediation~\cite{Lonzarich1}. Quantum oscillations that directly measure the effective quasiparticle mass are a crucial tool in revealing the divergent susceptibility signalling such a quantum critical point in diverse families of materials. A broader question pertains to whether such a potential quantum critical point underlying the superconducting dome is related to the observed enhancement in superconducting temperatures.

\section{Conclusion}

The measurement of quantum oscillations in the pnictide family of superconductors has greatly advanced our understanding of these materials. A Fermi surface comprising quasi-two dimensional hole and electron cylinders at the $\Gamma$ and $\tilde{M}$ points of the two-Fe ion Brillouin zone respectively in the overdoped paramagnetic side of the phase diagram evolves to a Fermi surface comprising small three-dimensional sections in the antiferromagnetic parent pnictide material via a potential quantum critical point where the quasiparticle effective mass is enhanced. The quasi-nesting wavevector $Q=\big(\pi/\tilde{a},\pi/\tilde{a}\big)$ that connects sections of the Fermi surface, and the potential quantum critical point under the superconducting dome appear to be common features of the pnictide with the cuprate family of superconductors. While enhanced antiferromagnetic correlations appear both in pnictide and cuprate superconductors, an open question still pertains to their relation to elevated superconducting temperatures.

\section{Acknowledgements}

SES acknowledges support from the Royal Society and King's College, University of Cambridge, and thanks A.~V.~Chubukov, M.~D.~Johannes,~G.~G.~Lonzarich, and~I.~I.~Mazin for useful discussions.



\section{Bibliography}\label{secbib}\index{bibliography}

\end{document}